   \documentclass[10pt,twocolumn,twoside]{IEEEtran}

\usepackage{amsmath,graphicx,cite}

\everymath{\displaystyle}
\title{Distributed Power Control with Partial Channel State Information: Performance Characterization and Design}

\author
  {Chao Zhang$^\dagger$, Samson Lasaulce$^\dagger$, Achal Agrawal$^*$, and Rapha$\ddot{\mathrm{e}}$l Visoz$^\ddagger$
	\thanks{$^\dagger$  L2S (CNRS-CentraleSupelec-Univ. Paris Sud), Gif-sur-Yvette, France.}
\thanks{$^*$ Mahindra Ecole Centrale, Hyderabad, India.}
\thanks{$^\ddagger$ Orange Labs, Issy-les-Moulineaux, France.}
\thanks{The material in this paper was partly presented  in \cite{Achal}.}}

\usepackage[mathcal]{euscript}
\usepackage{amssymb,amsmath}
\usepackage{algpseudocode}
\usepackage[]{algorithm2e}
\usepackage{verbatim}
\usepackage{graphicx}
\usepackage{longtable}
\usepackage{rotating}
\usepackage{multirow}
\usepackage{tabularx}
\usepackage{bigints}
\usepackage{tikz}
\usepackage{multirow} 
\usetikzlibrary{arrows,backgrounds,snakes}
\usepackage{amsmath,amstext,amsfonts,amssymb}
\usepackage{amsbsy}
\usepackage{amsthm}
\usepackage{diagbox}
\usepackage{mathabx}
\usepackage{color}
\usepackage{float}
\usetikzlibrary{arrows,shapes,positioning,shadows,trees}
\usetikzlibrary{decorations.pathreplacing}

\usepackage{float}
\usepackage{caption}
\usepackage{subcaption}

\usepackage{bbm}

\tikzset{
  basic/.style  = {draw, text width=2cm, drop shadow, font=\sffamily, rectangle},
  root/.style   = {basic, rounded corners=2pt, thin, align=center,
                   fill=green!30},
  level 2/.style = {basic, rounded corners=6pt, thin,align=center, fill=white!60,
                   text width=10em},
  level 3/.style = {basic, thin, align=left, fill=pink!60, text width=6.5em}
}

\definecolor{darkred}{rgb}{0.5,0.0,0.0}
\definecolor{darkblue}{rgb}{0.2,0.1,0.6}

\newcommand{\ol}{\overline}

\newcommand{\mc}{\mathcal}

\newtheorem{theorem}{Theorem}[section]

\newtheorem{proposition}[theorem]{Proposition}
\begin{document}
\maketitle

\begin{abstract} One of the goals of this paper is to contribute to finding distributed power control strategies which exploit efficiently the information available about the global channel state; it may be local or noisy. A suited way of measuring the global efficiency of a distributed power control scheme is to use the long-term utility region. First, we provide the utility region characterization for general utility functions when the channel state obeys an independent block fading law and the observation structure is memoryless. Second, the corresponding theorem is exploited to construct an iterative algorithm which provides one-shot power control strategies. The performance of the proposed algorithm is assessed for energy-efficient and spectrally efficient communications and shown to perform much better than state-of-the-art techniques, with the additional advantage of being applicable even in the presence of arbitrary observation structures such as those corresponding to noisy channel gain estimates.  

 \end{abstract}
   
\section{Introduction}

Many modern wireless networks tend to become distributed. This is already the case of Wifi networks which are distributed decision-wise; for example, each access point performs channel or band selection without the assistance of a central or coordinating node. As another example, small cells networks, which are envisioned to constitute one of the key components to implement the ambitious roadmap set for 5G networks \cite{hoydis-vtmag-2011}\cite{Bennis-2013}\cite{bastug-jwcn-2015}\cite{liu-tvt-2016} will need to be largely distributed; decentralization is one way of dealing with complexity and signalling issues induced by the large number of small base stations and mobile stations. In this paper, we consider wireless interference networks that are distributed both decision-wise and information-wise. More specifically, each transmitter has to perform a power control or, more generally, a radio resource allocation task by itself and by only having access to partial information of the network state. Being able to perform the corresponding power control task will be important since small cell networks will typically be interference networks.

When inspecting the literature on distributed power control (see e.g., \cite{book-zhuhan}\cite{lasaulce-book-2011}), it appears that the derived power control schemes are effectively distributed decision-wise and information-wise but almost always globally inefficient. A natural and important question arises. Is this because the considered power control scheme is not good enough or does it stem from intrinsic limitations such as limited information availability? To the authors' knowledge, this question has not been addressed formally. One of the goals of  this paper is precisely to provide a framework that allows one to derive the \textbf{limiting performance of power control in general multiuser channels} with partial information and therefore to be able to measure the efficiency of a given power control scheme. To reach this goal we resort to recent results that bridge the gap between decision theory and information theory \cite{ITW2015}. We exploit these results to characterize the limiting performance in terms of long-term utility region, each transmitter being assumed to have its own utility function. The performance characterization is then exploited in a constructive manner to determine power control strategies, and more specifically one-shot decision functions, which allow the transmitters to choose its power by only using the current available observation about the global channel state. The offline part of the proposed design is based on the knowledge of some channel statistics; this assumption becomes more and more realistic in wireless networks since more and more data should be  stored, processed and transmitted through the network. The practical interest in designing one-shot decision power control functions is very well motivated in the literature (see e.g., \cite{paul-icc}\cite{paul2014}); in particular, it allows the transmitter to take quick decisions, which do not generate extra delay (e.g., due to backhauling or non-direct inter-transmitter exchanges). Therefore, being able to design one-shot power control functions is of prime interest for small cells networks. 

To better assess the progress made on the design of one-shot or decision function compared to the state-of-the-art, let us explain the problem more formally. If Transmitter $i$ knows  ${g}_{ii}$ the channel gain of the link between Transmitter $i$ and Receiver $i$, the decision function writes under the form $f_i({g}_{ii})$. For example, in the pioneering work on energy-efficient power control \cite{goodman-2000} and more advanced works such as \cite{meshkati-poor}\cite{lasaulce-twc-2009}\cite{Haddad-2014}\cite{Bacci-TSP-2013}, the obtained distributed decision function is of the form of a channel inversion formula $f_i({g}_{ii}) = \frac{a}{{g}_{ii}}$, $a\geq 0$. Natural questions that are of practical importance and that have not been answered yet arise. How should the decision function be modified when only an estimate of the direct channel is available (say $\widehat{g}_{ii}$ instead of ${g}_{ii}$)? Which decision function should be chosen if another type of partial channel state information (CSI) is available? One of the contributions of the present paper is precisely to provide answers to these questions, namely, to be able to deal with the general case of \textbf{arbitrary partial information.} 

Elaborating further on the limitations of the state-of-the-art on the design of one-shot power control functions, existing works only deal with specific choices for the utility functions. In this respect, reference \cite{paul-icc} is a good representative of the corresponding literature and is particularly interesting. Therein, the authors show that using one-shot decision functions which are step functions may be optimal when the network utility function is chosen to be the Shannon sum-rate. But this result is again obtained by assuming a perfect knowledge for ${g}_{ii}$, which means that the above question still holds. However, there are other important related questions that appear. What would be the best decision function(s) in terms of global energy-efficiency e.g., when measured in terms of sum-energy-efficiency? More generally, what would be the best decision function(s) when an arbitrary utility function is considered for the network? The proposed framework allows one to answer such questions for \textbf{arbitrary utility functions} (namely, having the general form assumed in this paper).  

The main contributions of this paper can be summarized as follows:\\
$\bullet$ We provide the full characterization of the best achievable performance of distributed power control under arbitrary partial information and for an arbitrary utility function form. All available works either assume a particular partial information structure (e.g., perfect individual CSI) or a specific utility function (e.g., the sum-rate). \\
$\bullet$ We provide a general algorithm for determining one-shot power control functions which exploits the available information efficiently. The literature does not provide a systematic way of finding such functions, only ad hoc functions are proposed (such as thresholding functions).\\
$\bullet$ We conduct a thorough numerical analysis which relies on a simulation setting very similar to state-of-the art works.

This paper is structured as follows. In Sec. II, we provide the proposed general formulation of the problem of power control under partial information. In Sec. III, we derive the characterization of the best achievable performance of power control under partial information, which amounts to characterizing the long-term utility region. In Sec. IV, we propose one possible way of constructing good or efficient power control strategies under partial information. Sec. V corresponds to a detailed numerical analysis; to facilitate comparisons, several simulation scenarios have been chosen to be the same or almost the same as in the closest state-of-the-art papers. The paper is concluded by Sec. VI, which not only recaps some attractive features of our approach but also some of its limitations. 

\section{Problem statement}
\label{sec:problem-statement}

The wireless system under consideration comprises $K \geq 2$ pairs of interfering transmitters and receivers which can operate over $B\geq 1$ non-overlapping bands. The power Transmitter $i \in \{1,...,K\}$ allocates to band $b \in\{1,...,B\}$ is denoted by $a_i^b$, $a_i^b$ being subject to classical power limitations: $a_i^b \leq P_{\max}$ and $\sum_{b=1}^{B} a_i^b \leq P$, with $ P_{\max} \leq P$. Therefore, Transmitter $i$ has to adapt its \textit{power vector or action}: 
\begin{equation}
a_i = (a_i^1, ..., a_i^B)
\end{equation}
to the fluctuations of the channel gains of the different links between the transmitters and receivers, and to mitigate the interference. The channel gain of the link between Transmitter $i \in \{1,...,K\}$ and Receiver $j \in \{1,...,K\}$ for band $b \in \{1,...,B\}$ is denoted by $g_{ij}^b= |h_{ij}^b|^2 $. The \textit{global channel state} (also called non-controllable or nature action, hence the notation $a_0$) is then given by the following $K^2B-$dimensional vector which comprises all channel gains: 
\begin{equation}\label{eq:a_0-definition}
a_0 = (g_{11}^1,..., g_{11}^B,      g_{12}^1,..., g_{12}^B,   ..., g_{KK}^1,..., g_{KK}^B) 
\end{equation}
and is assumed to follow a given probability distribution which is denoted by $\rho_0$.

Transmitter $i$ can update its power vector $a_i $ from block to block. To update its power, each transmitter has a certain knowledge of the global channel state, which is called the \textit{partial information} available to Transmitter $i$ and is represented by the signal $s_i$. Before defining $s_i$, it has to be mentioned that for all the analytical and algorithmic results provided in this paper, the key quantities such as the power vector, the global channel state, and the partial information are assumed to be discrete (this assumption is discussed in detail at the end of the present section). This means that: $\forall i \in \{0,1,...,K\}, a_i \in \mathcal{A}_i$ with  $|\mathcal{A}_i| < \infty$ (the notation $|\cdot|$ stands for cardinality); $\forall i \in \{1,...,K\}, s_i \in \mathcal{S}_i$ with  $|\mathcal{S}_i| < \infty$. More specifically, the signal $s_i$ is assumed to be the output of a discrete memoryless channel whose transition probability is $\mathrm{P}(S_i=s_i | A_0 = a_0) =  \daleth_i(s_i| a_0)$ \cite{cover}, where $A_0$ and $S_i$  represent the random variables\footnote{To avoid any ambiguity where there is any, we use capital letters to refer to random processes or variables.} used to model the channel state variations and the partial information available to Transmitter $i$, respectively. The full or perfect global CSI at Transmitter $i$ corresponds to $s_i = a_0$. The case where only perfect individual CSI is available is given by $s_i = (g_{ii}^1, ...,g_{ii}^B)$. The signal $s_i$ may also be a noisy estimate of $(g_{ii}^1, ...,g_{ii}^B)$: $s_i = (\widehat{g}_{ii}^1, ...,\widehat{g}_{ii}^B)$. Note that in the numerical performance analysis, the proposed power control strategy is effectively computed by using discrete quantities but is tested over continuous channels (namely, channel gains correspond to realizations of complex Gaussian random variables). Indeed, each channel gain is assumed to obey a classical block-fading variation with independent realizations. This is a common assumption in the wireless literature (see e.g., \cite{caire-tit-1999}\cite{lau-tcom-2004}). The case of models like Gauss-Markov models is purposedly left as a relevant extension of the present work.

By denoting $t$ the block index, the purpose of Transmitter $i$ is therefore to tune the power vector $a_i(t)$ for block $t$ by exploiting its knowledge about the channel state, that is, the signal $s_i(t)$. More precisely, we assume that Transmitter knows $s_i$ not only at time $t$, but also the past realizations of it namely, $s_i(1), ..., s_i(t-1)$. The transmission is assumed to start at block $t=1$ and to stop at block $t=T$. In its general form, the \textit{power control strategy} of Transmitter $i$ is a sequence of functions which is denoted by $f_i = (f_{i,t})_{1\leq t \leq T}$ and defined by:
\begin{equation}\label{inf-struc}
\begin{array}{ccccc}
f_{i,t}  &:& \mathcal{S}_{i}^t       & \longrightarrow & \mathcal{A}_{i}\\
           &  &(s_i(1), s_i(2),...,s_i(t)) &\longmapsto&  a_i(t).
\end{array}
\end{equation}
Transmitter $i$ has to implement a power control strategy which aims at maximizing a certain performance metric called  \textit{utility function} in this paper and denoted by $u_i: \mc{A}_0 \times   \cdots \times \mc{A}_K \rightarrow \mathbb{R}$. 

The two main issues addressed in this paper are as follows. First, we characterize the achievable performance in terms of long-term utility region when the block or \textit{instantaneous utility} is a function of the form $u_i(a_0, a_1, ..., a_K)$. The \textit{long-term utility} of Transmitter $i$ is defined by:
\begin{equation}\label{eq:long-term-utility-def}
U_i(f_1,...,f_K) =   \lim_{T \rightarrow + \infty}  \displaystyle{\frac{1}{T} \sum_{t =1}^{T}} \mathbb{E} \big[ u_i(A_0(t), A_1(t), ..., A_K(t)) \big]
\end{equation}
whenever the above limit exists. The \textit{long-term utility region} therefore formally corresponds to all the points $(U_1, ...,U_K) \in \mathbb{R}^K$ that can be reached by considering all possible power control strategies as defined per (\ref{inf-struc}). The presence of the expectation operator is required in general (it can be omitted when a law of large numbers is available) since the channel is random and every power vector is a function of it. In general the channel is a random process $A_0(1),...,A_0(T)$ but since we will assume the channel gains to be i.i.d., the process can be represented by a single random variable $A_0$. The corresponding probability distribution is the \textit{global channel state distribution} (as already mentioned, it is denoted by $\rho_0$). The second issue we want to address in this paper is to determine power control strategies which only use the available local information while performing as well as possible in terms of a global utility e.g., in terms of sum-utility $\sum_{i=1}^K {U}_i$ with $u_i = \log\left(1+ \mathrm{SINR}_i \right)$ {(see \cite{david-2007})}. 

\textit{Remark 1.} Here, we would like to provide our motivations for assuming discrete quantities in the analytical part and the algorithmic part of this paper. First, as far as the limiting performance or achievable utility region characterization is concerned, one can easily invoke standard arguments (see \cite{cover}) to show that the continuous case follows from the discrete case (the main change is to replace sums with integrals in the equations characterizing the limiting performance). Second, there exist real systems where the transmit power has to be discrete (see e.g., \cite{sesia-book-2009}). Quite remarkably, imposing the transmitters to use a reduced action space may be beneficial both for the network and individual performance; simulations provided in this paper show that the very interesting result obtained in \cite{david-2007} is in fact more general; in \cite{david-2007}, the authors show that binary power control may be optimal or generate a very small performance loss compared to the continuous case. In this respect,  the authors have shown in \cite{paul-icc} that using one-shot decision functions which are step functions may be optimal when the utility function is chosen to be the Shannon sum-rate. One of the important contributions of the present work can be seen as a generalization of such a result to \textbf{arbitrary utility functions}. Third, concerning channel gains, the proposed approach is to assume that actual channel gains are continuous but the algorithm used to find the power control policy is using quantized values. This can be verified to induce a typically small performance loss as soon as the cardinality of the set of channel gains exceeds $8-10$. Otherwise, the proposed algorithm has to be adapted to the continuous case, which is left  as a possible extension of this paper.

\textit{Remark 2.} Considering the long-term utility as defined by (\ref{eq:long-term-utility-def}) is very relevant regarding to the performance analysis methodology that is used in the literature of radio resource allocation. Indeed, typically, a policy is determined for a given realization of the channel and its average performance is assessed by performing Monte Carlo simulations. Here, we directly consider the ultimate (average) performance criterion and not just the instantaneous performance. The offline part of the proposed design to determine a policy (say a function) which provides good average performance is based on the knowledge of some statistics on the channel; this assumption becomes more and more realistic in wireless networks since more and more data become available, are stored, and processed. To be implemented online the policy or function only needs its own arguments i.e., the information available online (e.g., an estimate of a given channel gain). At last, notice that if the channel statistics vary over time, it is always possible to use a sliding time-window to adapt to those possible variations. These secondary effects are not studied here.

\section{Limiting performance characterization of distributed power control with partial information}\label{sec:utility-region}

While many power control schemes using partial CSI are available in the literature, very often it is not possible to know whether the available information is exploited optimally by the considered power control scheme. While the problem of optimality is in general a very important and challenging problem, it turns out to be solvable in important scenarios such as the scenario under investigation in this paper. Indeed, an important message of the present work is that, under the made assumptions, information theory tools can be used to fully characterize the limiting theoretical performance of the power control strategies. The two key assumptions which are made for this are as follows: (i) The channel state $a_0(t)$ is i.i.d.; (ii) The observation structure which defines the partial observation $s_i$ is memoryless. Assuming (i) and (ii), the theorem which is given a bit further provides the utility region characterization for any power control problem under the form defined in the previous section. 

Define the vector $a=(a_0, a_1,...,a_K) \in \mc{A}$. To better understand the characterization of the limiting performance of distributed power control which is given through Theorem \ref{thm}, it is useful to observe that the sum over time $\frac{1}{T} \sum_{t=1}^T u_i(a(t))$ can always be re-written under the form $\sum_{a \in \mc{A}} P^{(T)}(a) u_i(a) $ where $P^{(T)}(a)$ is the number of times that the realization $a$ appears in the sequence $a(1),...,a(T)$ divided by $T$; $P^{(T)}(a)$ is by construction a probability. Therefore, the performance of any power control policy can be fully characterized in terms of achievable joint frequencies of probabilities over $\mc{A}$. This explains why probabilities appear in Theorem \ref{thm}. Obviously, not all joint probabilities of the unit simplex are achievable. For instance, Transmitter $i$ would not be able to maximize (for a given block $t$) the utility $u_i(a_0(t), a_1(t), ..., a_K(t))$. Indeed, the corresponding distribution would be $P_{A_0 A_1...A_K}(a) = P_{A_1...A_K|A_0}(a_1,...,a_K|a_0)  P_{A_0}(a_0) = \mathbbm{1}_{(a_1,...,a_K) \in \phi(a_0)}   \rho_0(a_0)$ where $\phi$ is the set-valued function corresponds to the argmax of $u_i$ for the example here ($\mathbbm{1}_{\cdot}$ stand for the indicator function). This distribution is trivially non-achievable in general since it would mean that Transmitter $i$ has full access to the global CSI $a_0(t)$ and is able to control all the decision variables $(a_1(t), ..., a_K(t))$. 

For the sake of clarity, we will use the following notations: $s=(s_1,...,s_K) \in \mc{S}$; $\daleth$ stands for the conditional probability $\mathrm{P}_{S|A_0}$, $S=(S_1, ...,S_K)$ being the random variable used to model the vector of individual signals available to the transmitters; $V$ is an \textit{auxiliary variable} as used in coding theorems. Auxiliary variables are commonly used to state coding theorems, see for instance the work on the broadcast channel \cite{cover} or \cite{dikstein-tit-2016} for a more recent reference. Its operational meaning in our context will be provided a little further. The notation $\Delta_{K}$ will refer to the \textit{unit simplex} of dimension $K-1$: 
\begin{equation}
\Delta_{K} =
\left\{(x_1,...,x_K) \in \mathbb{R}^K:  \forall i \in \{1,...,K\}, x_i \geq 0;      
\displaystyle{\sum_{i=1}^K} x_i = 1 \right\}.
\end{equation}
At last, we define the following notations:  the function $w_{\lambda}  = \displaystyle{\sum_{i=1}^K} \lambda_i u_i$ represents the \textit{weighted utility} with $\lambda  =(\lambda_1,...,\lambda_K) \in \Delta_K$; the function $W_{\lambda}$ represents the expected version of the function $w_{\lambda} $ i.e., $W_{\lambda} = \mathbb{E}(w_{\lambda})$. The expected weighted utility $W_{\lambda}$ can always be written as:
\begin{equation}
\begin{array}{ccl}
W_{\lambda} & = & \sum_{a} P_A(a) w_{\lambda}(a) \\
                     & =  & \sum_{a,s,v} P_{A S V}(a, s, v)   w_{\lambda}(a) 
\end{array}.
\end{equation}
The point is that not all joint probabilities $P_{A S V}$ can be realized. For a given observation structure (given by $\daleth$), only some probabilities can be obtained by considering all the possible power control policies. Only probabilities which write as in Theorem \ref{thm} are achievable. 

\begin{theorem} When $T \rightarrow + \infty$, the region of achievable long-term utilities is given by
\begin{equation}
\mathcal{U} = \bigg\{   (U_1,...,U_K) \in \mathbb{R}^K: 
U_i =  \displaystyle{\sum_{a,s,v}}  Q(a,s,v) u_i(a), Q \in \mathcal{Q} \bigg\} 
\label{eq:Ubar}
\end{equation}
where $\mathcal{Q}$ is defined by
\begin{equation}
\begin{array}{l}
\mathcal{Q}  =  \Big\{Q \in \Delta_N:  Q(a,s,v) = \rho_0(a_0) \daleth(s|a_0) P_V(v)\times  \\  

\quad\quad\quad\quad\quad\quad\quad\quad\quad\quad\quad\quad \displaystyle{\prod_{i=1}^K } P_{A_i|S_i,V} (a_i|s_i,v)  \Big\}, 
\end{array}
\label{eq:Qset}
\end{equation}
$P_V$ is the distribution of some auxiliary variables $V \in \mathcal{V}$ verifying the Markov chain $V- (A_0,A_1,...,A_K) - (S_1, ...,S_K)$, and $N= |\mathcal{V}| \prod_{k=0}^K |\mc{A}_k| \prod_{\ell=1}^K |\mc{S}_\ell| $. The long-term utility region can be entirely covered by imposing the range for the cardinality of the set $V \in \mathcal{V}$  to be bounded as follows:
\begin{equation}
|\mathcal{V}|\leq |\mathcal{A}|\cdot{|\mathcal{S}|}-1,
\end{equation}
where $|\mathcal{A}|=\prod_{k=1}^K |\mc{A}_k|$ and $|\mathcal{S}|=\prod_{\ell=1}^K |\mc{S}_\ell|$.
\label{thm}
\end{theorem}
\textbf{Proof:} The proof comprises three steps. First, we prove that the long-term utility $U_i$ is the linear image of the average joint probability distribution over $\mc{A}$, which proves that characterizing the region of long-term utilities amounts to characterizing the achievable joint probabilities. Second, we apply a theorem to determine the set of achievable joint probabilities. Third, we exploit the support lemma \cite{IT1975}\cite{IT1976} to derive the upper bound on the cardinality of $\mc{V}$.

\textbf{Step 1}. Linear relation between $U_i$ and $\lim_{T \rightarrow + \infty}  \frac{1}{T} \sum_{t =1}^{T} P_{t}(a_0,..., a_K)  $.\\ 
We show that the power control strategies of the different users $f_1,...,f_K$ intervene in the long-term utility only through the joint probability over $\mc{A}_0 \times ...\times \mc{A}_K$. Therefore, characterizing the long-term utility region is equivalent to characterizing the set of achievable or implementable joint probability distributions. We have that:

\begin{eqnarray}
U_i(f_1,...,f_K) &&\\
= \displaystyle{  \lim_{T \rightarrow + \infty} } \displaystyle{\frac{1}{T} \sum_{t =1}^{T}} \mathbb{E} \left[ u_i(A_0(t), A_1(t), ..., A_K(t)) \right]&&\\
=  \displaystyle{\lim_{T \rightarrow + \infty}}  \displaystyle{\frac{1}{T} \sum_{t =1}^{T}} \displaystyle{\sum_{a_0,...,a_K}} P_{t}(a_0,..., a_K) u_i(a_0,...a_K)&& \\
=  \displaystyle{\sum_{a_0,...,a_K}} u_i(a_0,...,a_K)  \displaystyle{\lim_{T \rightarrow + \infty}}  \displaystyle{\frac{1}{T} \sum_{t =1}^{T}} P_{t}(a_0,..., a_K) &&
\end{eqnarray}
where $P_{t}(a_0,..., a_K) $ is the joint probability distribution induced by the power control strategy profile $f_1,...,f_K$ at time $t$. Again, we denote the random process $A_i(t)$ by capital letters to distinguish it from its realization, denoted by $a_i$.  Therefore, a utility level $\mu_i $ is achievable if and only if it can be written as
\begin{equation}
\mu_i = \displaystyle{\sum_{a_0,a_1,...,a_K}} Q( a_0,a_1,...,a_K ) u_i(a_0,a_1,...,a_K)
\end{equation}
 and there exists a power control strategy profile $(f_1,...,f_K)$ such that
\begin{equation}
\lim_{T\rightarrow \infty} \frac{1}{T} \sum_{t =1}^{T} P_{t}(a_0,..., a_K) = Q(a_0,..., a_K).
\end{equation} 


\textbf{Step 2.} Set of achievable joint distributions.\\
By using the coding theorem of \cite{ITW2015} one has that a joint probability distribution $Q(a_0,a_1,...,a_K)$ is implementable if and only if it factorizes as:
\begin{equation}
Q(a) =\rho_0(a_0)   \sum_{s,v} \daleth(s|a_0) P_V(v) \displaystyle{\prod_{i=1}^K }
P_{A_i|S_i,V}(a_i|s_i,v)  
\label{eq:prob_general}
\end{equation}
where $V$ is any random variable which verifies the Markov chain $V- (A_0,A_1,...,A_K) - (S_1, ...,S_K)$ and whose cardinality is upper bounded as follows.

\textbf{Step 3.} Bounding the cardinality of $\mc{V}$.\\
The proof is based on the following lemma.\\
\textit{Support Lemma.}  Let $\mathcal{X}$ be a finite set and $\mathcal{V}$ be an arbitrary set. Let $\mathcal{P}$ be a connected compact subset of probability distributions on $\mathcal{X}$ and $p(x|v)\in \mathcal{P}$, indexed by $v\in \mathcal{V}$, be a collection of (conditional) pmfs on $\mathcal{X}$. Suppose that $\eta_j(\pi)$, $j \in \{1,...,d\}$, are real-valued continuous functions of $\pi\in \mathcal{P}$. Then for every $V\sim F(v)$ defined on $\mathcal{V}$, there exists a random variable $V'\sim p(v^{\prime})$ with $|\mathcal{V^{\prime}}|\leq d$ and a collection of conditional probability distributions $p(x|v')\in\mathcal{P}$, indexed by $v'\in \mathcal{V}'$, such that for $j \in \{1,...,d\},$
\begin{equation}
\int_{v \in \mathcal{V}}  \eta_j(p(x|v)) \mathrm{d} F(v)=\sum_{v'\in\mathcal{V}^{\prime}} \eta_j(p(x|v^{\prime}))p(v^{\prime}).
\end{equation}
Now let us show how this lemma is used to bound the cardinality of auxiliary random variables. Suppose $\mathcal{X}=\mathcal{A} \times \mathcal{S}$, which refers to the joint action and joint state (observation) profiles. The corresponding $\mathcal{P}$ will be a connected compact subset of probability distributions on $\mathcal{A} \times \mathcal{S}$ and $p(a_1,...,a_K,s_1,...,s_K|v)\in \mathcal{P}$, indexed by $v\in \mathcal{V}$, be a collection of (conditional) p.m.f. on $\mathcal{A} \times\mathcal{S}$. 
Note that the product distribution $\displaystyle{\prod_{i=1}^K } P_{A_i|S_i,V}(p_i|s_i,v)$ constitutes a special form of the general probability form $P_{A_1,...,A_K|S_1,...,S_K,V}(a_1,...,a_K|s_1,...,s_K,v)$ and the general probability form can be rewritten as:
\begin{equation}
\begin{split}
&P_{A_1,...,A_K|S_1,...,S_K,V}(a_1,...,a_K|s_1,...,s_K,v)\\
=&\frac{P_{A_1,...,A_K,S_1,...,S_K|V}(a_1,...,a_K,s_1,...,s_K|v)}{P_{S_1,...,S_K|V}(s_1,...,s_K|v)}.
\end{split}
\end{equation}
Hence,  $\displaystyle{\prod_{i=1}^K } P_{A_i|S_i,V}(a_i|s_i,v)$ can be expressed by $\pi\in\mathcal{P}$. Denoting by $j_K$ the ratio $j$ to $K$, consider the following $|\mathcal{A}|-1$ continuous functions on $\mathcal{P}$:
\begin{equation}
\eta_j(\pi)=\frac{\pi(j)}{\displaystyle\sum_{i=j_K+1}^{i=j_K+K}\pi({i})}\,\,\,\,\,\,\,j\in \{1,...,|\mathcal{A}|\times|\mathcal{S}|-1\}.
\end{equation}
Clearly, these $|\mathcal{A}|\times |\mathcal{S}|-1$ functions are continuous.  According to the support lemma, for every $V\sim P_{V}(v)$ defined on $\mathcal{V}$, for the distribution $Q(a)$, there exist a $V^{\prime}\sim P_{V^{\prime}}(v^{\prime})$ with $|\mathcal{V}^{\prime}|\leq |\mathcal{A}|\cdot|\mathcal{S}|-1$ such that
\begin{equation}
\begin{split}
Q(a) &=\rho_0(a_0)  \sum_{s,v} \daleth(s|a_0)  P_V(v) \displaystyle{\prod_{i=1}^K }
P_{A_i|S_i,V}(a_i|s_i,v)  \\
&=\rho_0(a_0)  \sum_{s,v^{\prime}} \daleth(s|a_0)  P_{V^{\prime}}(v^{\prime}) \displaystyle{\prod_{i=1}^K }
P_{A_i|S_i,V^{\prime}}(a_i|s_i,v^{\prime}).  \\
\end{split} 
\label{eq:prob_3}
\end{equation}

To better understand Theorem III.1 and its proof, let us comment on it in detail.

The first comment which can be made is that the long-term utility region characterization relies on the use of an auxiliary random variable $V$. The presence of such variables is very common in coding theorems. For example, the capacity region of degraded broadcast channels is parameterized by auxiliary variables; for one transmitter and two receivers, only one auxiliary variable suffices. In the latter case, the auxiliary variable can be interpreted for instance as a degree of freedom the transmitter has for allocating the available resource between the two receivers \cite{cover}. In general, auxiliary random variables have to be considered as parameters which allow one to describe a set of points and therefore constitute, before all, a mathematical tool. In the context of this paper, from the design point of view, $V$ may be interpreted as a coordination random variable or a lottery which allows one to generate a coordination key; this key would be generated offline, which means incurring no loss in terms of performance. By the way of an example, consider a single-band interference channel with two transmitters and two receivers and let us describe one possible way of implementing the suggested coordination mechanism. The idea is to exchange a coordination key offline and which consists of a sequence of realizations  $v(1),..., v(T)$ of a (Bernoulli) binary random variable: $V \sim \mathcal{B}(\tau)$, $\tau \in [0,1]$. Then, online, a possible rule for the transmitters might be as follows: if $v(t)=1$, Transmitter $1$ transmits and    
if $v(t)=0$, Transmitter $2$ transmits. We see that in this particular example, $V$ would act as a time-sharing variable which would allow to manage interference even if the transmitters have no knowledge at all about the channel (i.e., $s_i = const.$). Then, by optimizing the Bernoulli probability $\tau$, one can obtain better performance than transmitting at full (or constant) power. Note that the full power operation point would be obtained by applying the iterative water-filling algorithm (IWFA) (see e.g. \cite{yu-2002}\cite{scutari-tsp-2009})  to a single-band interference network where each Transmitter $i$ wants to maximize its utility $u_i = \log(1+\mathrm{SINR}_i)$, $\mathrm{SINR}_i$ being the SINR at Receiver $i$. In fact, the usefulness of such a costless coordination mechanism is even more apparent when QoS constraints are accounted for, as done in one scenario of the numerical performance analysis.

The second comment we will make here is that the power control strategy only intervenes in the long-term utility through its average behavior i.e., in terms of conditional probability $P_{A_i|S_i,V}$, that is, the (conditional) frequency at which a given power vector $a_i$ is used. Optimality of a given power control strategy under partial information is only related to the frequencies at which the possible transmit power levels are used.
 
The third comment concerns the alphabet $V$ lies in, namely $\mathcal{V}$. Indeed, it is possible to cover all the feasible utility region by choosing appropriately the possible range for $|\mathcal{V}|$. Indeed, in general, to cover all the feasible utility region, the range for $|\mathcal{V}|$ has to vary in an interval which is specified in Theorem III.1. Considering larger values for $|\mathcal{V}|$ would not bring any performance gain. This argument is especially useful for avoiding useless extra computational burden when computing the long-term utility region.

The last comment concerns the derived result itself. The utility region can be seen as a counterpart of the notion of capacity region. Indeed, a general power control strategy is a sequence of functions and each of these functions maps a sequence of realizations to a symbol. Therefore, mathematically speaking, a power control strategy is a code. Obviously, this code has not the same role as the source or channel codes which operate at the bit or symbol level and not at the data block level. This connection has been established for the first time in \cite{larrousse-isit-2013} and is not really exploited here because the available information is causal. For more details about this interpretation see \cite{larrousse-isit-2013}, \cite{larrousse-tit-2018}, \cite{dikstein-tit-2015}.

\section{Proposed power control strategies}
\label{sec:pc-strategies}

The knowledge of the long-term utility region is a very precious tool since it allows the efficiency of any distributed power control scheme to be assessed. But the problem of designing efficient distributed power control schemes remains open: There is no general recipe to find power control schemes which allow one to operate arbitrarily close to a point of the utility region established through Theorem III.1. This is reminiscent to the problem of designing multiuser channel codes knowing the capacity region. To obtain effectively implementable strategies, we make here particular choices. These choices are well justified a posteriori by numerical comparisons with state-of-the-art strategies. We choose to focus on memoryless and stationary power control strategies. Our motivation is twofold. First, it effectively allows one to design distributed power control schemes that use the available information. Second, one-shot strategies are very useful in practice e.g., in small cells networks. A strategy is \textit{memoryless} in the sense that it does not exploit the past realizations of the signal $s_i$; it is therefore a sequence of functions which writes as $f_{i,t}(s_i(t))$. A strategy is \textit{stationary} in the sense that the function $f_{i,t}$ does not depend on time, which ultimately means that a power control strategy boils down to a single function say $\overline{f}_i$; the latter function will be referred to as a \textit{decision function}. In fact, considering that the power level, vector, or matrix of a transmitter only depends on the current realization of the channel, and this in a stationary manner, is a very common and practical scenario in the wireless literature \cite{paul-icc}. As advocated by recent works  (see e.g., \cite{paul-spawc} for the MIMO case), the problem of finding one-shot decision functions with partial information and which perform well in terms of global performance is still a challenging problem. Remarkably, one of our observations is that Theorem III.1 can be exploited in a constructive way that is, it can be exploited to find good decision functions. This is precisely the purpose of this section. 

The key observation we make is as follows. Since the long-term utility region is convex (this is ready since $V$ can be used as a time-sharing variables over any $K-$uplet of distributed strategies), one can therefore operate on its Pareto frontier by maximizing the expected weighted utility, which is denoted by $W_{\lambda} = \mathbb{E}(w_{\lambda})$, with $w_{\lambda}  = \displaystyle{\sum_{i=1}^K} \lambda_i u_i$ and $\lambda  =(\lambda_1,...,\lambda_K) \in \Delta_K$. According to Theorem \ref{thm}, the achievable expected weighted utilities write as $W_{\lambda} = \sum_{a,s,v} Q(a,s,v) w_{\lambda}(a)$ where $Q \in \mc{Q}$. The functional $W_{\lambda}$ is thus a multilinear function of its arguments which are conditional probability distributions  $P_{A_1|S_1,V}, ...,  P_{A_K|S_K,V}, P_V$. Its maximum points are on the vertices of the unit simplex \cite{bilinear}. This means that, in the absence of QoS constraints, choosing $V=\mathrm{const}$ allows one to reach the corner points of the utility region and optimal conditional probabilities $P_{A_i|S_i,V}$ boil down to functions under the form $a_i = f_i(s_i)$. Therefore, corner points can be reached by using stationary and memoryless strategies. 

The key idea is to solve the corresponding optimization problem to determine these functions and use them as candidates for power control decision functions. This is why we will denote these functions by $\overline{f}_i$,  $i \in \{1,...,K\}$. Finding a low-complexity numerical technique to determine the optimal functions is left as a relevant extension of the present work. Instead, here we propose a suboptimal optimization technique which relies on the use of the sequential best-response dynamics (see e.g., \cite{lasaulce-book-2011}\cite{tirole}), which has the merit to be implementable from the computational standpoint. To apply the sequential best-response dynamics to $W_{\lambda}$, we rewrite it by isolating the sum w.r.t. $s_i$ i.e., the observation of Transmitter $i$:

\begin{eqnarray}
W_{\lambda} &=& \displaystyle\sum_{a_0,s} \rho_0({a}_0)  \daleth(s | {a}_0 ) w_{\lambda}(a_0,  \overline{f}_1(s_1),...,\overline{f}_K(s_K)  ) \nonumber\\
&=&\displaystyle\sum_{{a}_0,s_i}\big[  \rho_0({a}_0)  \daleth_i(s_i|a_0) \displaystyle\sum_{s_{-i}} \daleth_{-i}(s_{-i}|a_0)\times   \nonumber\\  &&\quad\quad\left.w_{\lambda}   (a_0, \overline{f}_1(s_1),...,\overline{f}_K(s_K))\right]
\label{eq:pc_general}
\end{eqnarray}
where: $s_{-i} = (s_1, ...,s_{i-1},s_{i+1}, ...,s_K )$ represents the vector comprising all observations of the transmitters other than Transmitter $i$; the condition probability $\daleth_{-i}$ is given by
\begin{equation}
\daleth_{-i}(s_{-i}|a_0)
= \displaystyle{\sum_{s_i} \daleth(s | a_0)}. 
\end{equation}

To describe the proposed iterative algorithm, it is convenient to introduce the following auxiliary quantity:
\begin{equation}
\label{theta}
\begin{array}{l}
{\omega_i(s_i,a_i)} = \displaystyle\sum_{{a}_0}  \big[   \rho_0({a}_0) \daleth_i(s_i|a_0) \displaystyle\sum_{s_{-i}}  \daleth_{-i}(s_{-i}|a_0) \times    \\  \left.   w_{\lambda}(a_0,       
 \overline{f}_1(s_1),...   ,  \overline{f}_{i-1}(s_{i-1}), a_i, \overline{f}_{i+1}(s_{i+1}),...,\overline{f}_K(s_K))) \right].
\end{array}
\end{equation}
The sequential best-response dynamics procedure consists in updating one variable at a time, the others being fixed, and proceed in a round-robin manner. Here, the variables are the decision functions. By assuming the knowledge of the utility function $w_{\lambda}$, the alphabets $\mc{A}_0, \mc{A}_1, ..., \mc{A}_K$, $\mc{S}_1,...,
\mc{S}_K$, the probability distribution of the channel $\rho_0$, the distribution of the observed signals $\daleth$, and an initial choice for the decision functions $\overline{f}_1^{\mathrm{init}}, ..., \overline{f}_K^{\mathrm{init}}$ Algorithm 1 can be implemented offline. The proposed algorithm would typically be implemented offline, whereas the obtained decision functions are designed to be exploited online. Algorithm 1 can thus either be used by a designer who performs a performance analysis based on propagation scenario he can define and know perfectly in terms of statistics or by a terminal of the network which is able to collect enough data to estimate the needed statistics. Even though the decision function determination operation requires the knowledge of the different alphabets, the channel statistics, the observation signal statistics, and the initial decision functions, it is essential to notice that \textbf{Transmitter $i$ only needs $s_i$ and possibly $v$} to tune (online) its power vector.  

\begin{algorithm}[]
\label{algo1}
    \SetKwInOut{Input}{inputs}
    \SetKwInOut{Output}{outputs}
    \SetKwRepeat{Do}{do}{while}%
\caption{Proposed decentralized algorithm for finding decision functions for the transmitters}
\Input{$\mc{A}_0$; $\forall i \in \{1,...,K\}, \ \left( \mathcal{A}_i,  \mathcal{S}_i \right)$\newline
 $w_{\lambda}, \ \rho_0, \ \daleth$\newline
 $\forall i \in \{1,...,K\}, \   \overline{f}_i^{\mathrm{init}}$} 
\Output{$\forall i \in \{1,...,K\}, \  \overline{f}^{\star}_i$}
~\\
Initialization: $\overline{f}_i=\overline{f}_i^{\mathrm{init}}$, $\overline{f}_i^{\mathrm{OLD}}=\frac{1}{2}\overline{f}_i^{\mathrm{init}}$,\, $\mathrm{iter}=1$\\
~\\
\While{$\exists i:$ $\|\overline{f}_i^{\mathrm{OLD}} - \overline{f}_i\|_2\geq \epsilon \,$ AND $\, \mathrm{iter} \leq \mathrm{iter}_{\max}$ $\qquad$ }{
~\\
\ForEach{$i \in \{1,\hdots,K\}$}{%
~$\overline{f}_i^{\mathrm{OLD}}=\overline{f}_i$;\\
\ForEach{$s_i \in \mathcal{S}_i$}{%
~\\
$\overline{f}_i(s_i)= \arg \max_{a_i} \omega_{i}(s_i,a_i)$ (Defined by (\ref{theta})) :
}
}
$\mathrm{iter} = \mathrm{iter}+1$\;
}
Final update: $\forall i \in \{1,...,K\}, \  \overline{f}^{\star}_i  =  \overline{f}_i$
\end{algorithm}

A classical issue is to know whether this iterative algorithm converges. For clarity, we state the following convergence result under the form of a proposition.

\begin{proposition}
\label{prop2}
Algorithm 1 always converges.
\end{proposition}

\textbf{Proof:} The result follows since the functional $W_{\lambda}(\ol{f}_1, ..., \ol{f}_K)$ is non-decreasing with the iteration index and is bounded. $\ \blacksquare$

Obviously, there is no guarantee for global optimality and only local maximum points for $W_{\lambda}$ are reached in general by implementing Algorithm 1. Quantifying the optimality gap is known to be a non-trivial issue related to the problem of determining a tight bound of the price of anarchy \cite{lasaulce-book-2011}\cite{papadimitriou}. Two comments can be made. First, if the algorithm is initialized by the best state-of-the-art decision functions, then it will lead to new decision functions which perform at least as well as the initial functions. Second, many simulations performed for a large variety of scenarios have shown that the optimality gap seems to be relatively small for classical utility functions used in the power control literature.

\textit{Remark 3.} To take QoS constraints into account (as done in related works such as  \cite{yates},  \cite{fu}), the maximization of $W_{\lambda}$ should be performed under constraints written in terms of expected constraint functions. In this situation, optimal solutions may not be just functions but conditional probabilities that lie in the interior of the set of probability distributions. This would mean that a transmitter has to randomize over its possible actions and a non-trivial optimal lottery has to be determined for $V$. In the numerical part, a simple scenario which corresponds to this situation is treated. Therein, the optimal $V$ corresponds to time-sharing over memoryless and stationary strategies. 

\textit{Remark 4.} The computational complexity of Algorithm 1 is $O(I \times  |\mathcal{A}_0|\times |\mathcal{S}|\times\sum_{k=1}^{K} |\mathcal{A}_k|)$ where $I$ is the number of iterations of the algorithm. As a rule of thumb, it is known that the total number of iterations for convergence (say up to a few percents) is typically $3K$ or $4K$, $K$ being the number of transmitters (see e.g., \cite{lasaulce-book-2011}). In practice, $K$ is the number of transmitters that effectively interfere (that is, using the same band at the same time in the same area). It is not very large in practice, say between $2$ and $10$. The complexity might be decreased further e.g., by accounting for the specific form of the utility function, its properties (such as convexity), and by using approximations. But, as mentioned before, the problem of designing low-complexity procedures to determine optimal decision functions is left as an extension of the present work.

%
%
%

\section{Numerical performance analysis}
\label{sec:numerical-analysis}
%
%
%
%
%
%

In this section, unless explicitly mentioned otherwise, our attention will be dedicated to energy-efficient power control. The reason for this is twofold. First, the problem of rate-efficient power control (for which it is typically assumed that $u_i = \log(1+ \mathrm{SNR}_{i})$) has been largely addressed in the literature, albeit almost always in presence of perfect individual CSI. Note that individual CSI may be easily acquired e.g., when each transmitter sends a pilot or training sequence to its intended receiver. Then, Receiver $i$ can estimate $g_{ii}$ and feedback it to Transmitter $i$. Second, designing energy-efficient communications is becoming a more and more important issue in real wireless systems. The assumed sum-utility function is the one used e.g, in \cite{betz2008}\cite{vineeth-2012}\cite{Buzzi-survey-2016} and the references therein:
\begin{equation}
w^{\mathrm{EE}}(a_0,a_1, ...,a_K) = \sum_{i=1}^{K}  \frac{ R_0 \times \psi(\mathrm{SINR}_i)}{a_i+P_0} 
\end{equation}
where $R_0$ is the {raw data rate (in bit/s) } and $\psi$ is a function which represents the net data rate. The function $\psi$ might represent the packet success rate (see e.g., \cite{meshkati-poor} where $\psi(x) = (1-e^{-x})^M$, $M\geq1$ being the packet length), the complementary of the outage probability (see e.g., \cite{Veronica2011} where $\psi(x) = e^{-\frac{c}{x}}$, $c>0$ being a constant related to spectral efficiency), or the Shannon spectral efficiency (see e.g., \cite{verdu-1990}\cite{Veronica-2009}\cite{Zappone-TWC-2013} where $\psi(x) = \log(1+x)$).  The raw data rate is  the same as \cite{Zappone-TWC-2013}, i.e., $R_0=1$ Mbit/s. At last, the constant $P_0$ represents the power consumed by the transmitter when the radiated power is zero. For instance, in \cite{betz2008} it may represent the computation power, the circuit power, or the base station power consumption as in \cite{vineeth-2012}\cite{jorswieck-2012}.  

{To implement Algorithm 1, quantized channel gains are used. To obtain the channel gain alphabet $\mathcal{G}$ in which each channel gain $g_{ij}^b = |h_{ij}^b|$ lies, we apply a maximum entropy quantizer \cite{chao-twc} to the modulus of $h_{ij}^b$, the real and imaginary parts of $h_{ij}^b$ being Rayleigh distributed. Also we will assume that $V=const$. At the end of this section, however, we shall provide numerical results  to get a bit more insights about the choice of the auxiliary variable $V$, knowing that the proper design of the coordination key is an interesting issue which is left as an extension of this paper.  }

\subsection{Influence of the channel estimation quality of the individual CSI on the shape of decision functions}

{In this subsection, the cardinality of $\mathcal{G}$ is set to $15$: $|\mathcal{G}|=15$. For the ease of exposition, we shall choose the reference scenario given by the following choices for the model parameters: $K=2$, $B=1$, $\sigma^2=10$ mW, $P_{\max}=100$ mW, $\mathbb{E}(g_{ii})=1$, for $j\neq i$, $10 \log_{10}\left( \frac{\mathbb{E}(g_{ii})}{\mathbb{E}(g_{ji})}   \right) =5$ dB, $\mathcal{A}_i = \left\{0, \frac{P_{\max}}{|\mathcal{P}|-1}, \frac{2 P_{\max}}{|\mathcal{P}|-1}, ...,  P_{\max}  \right\}$ for the power level alphabet with $|\mathcal{A}_i|=\mathcal{P}|=75$, and $R_0=1$ Mbit/s (or $1$ MHz). We assume that Transmitter $i$, $i \in \{1,...,K\}$ has some imperfect knowledge about the individual CSI i.e., $s_i = \widehat{g}_{ii}$.} To obtain the channel gain estimate $\widehat{g}_{ii}$ we consider a noisy version of the actual (continuous) channel gain with $\widetilde{g}_{ii} = {g}_{ii} + z_{i}$ ($z_{i}$ being an AWGN) and apply the aforementioned quantization operation to obtain $\widehat{g}_{ii}$. This defines a certain estimation SNR (ESNR) which is given by:
\begin{equation}\label{eq:ESNR}
\mathrm{ESNR}_{i}= \frac{\mathbb{E}[ {g}_{ii}^2]   }{\mathbb{E}[ ( \widehat{g}_{ii} -  {g}_{ii})^2]}.
\end{equation}

\begin{figure}[h]
   \begin{center}
        \includegraphics[width=.48\textwidth]{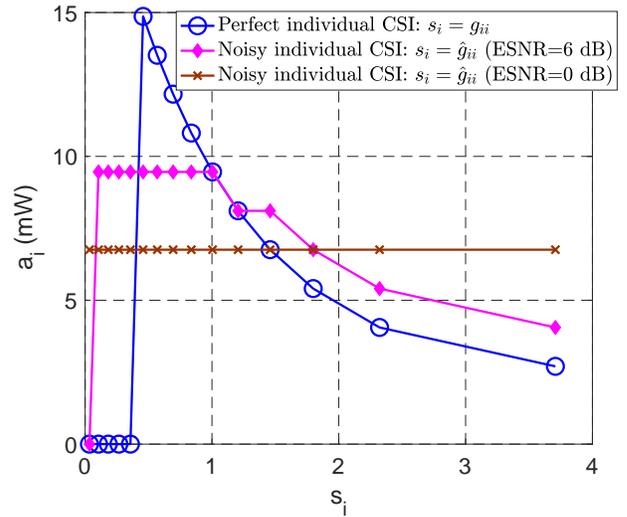}
    \end{center}
  \caption{When the estimation noise level increases, it is seen that at optimum, only some of the available transmit power levels are exploited to maximize sum-energy-efficiency (namely, by using Algorithm 1) here with $\psi(x) = e^{-\frac{c}{x}}$ and $P_0=0$. In connection with the literature  on learning we refer to this effect as a cooling effect.}
   \label{fig:dec_fun}
\end{figure}

\begin{figure}[h]
   \begin{center}
        \includegraphics[width=.48\textwidth]{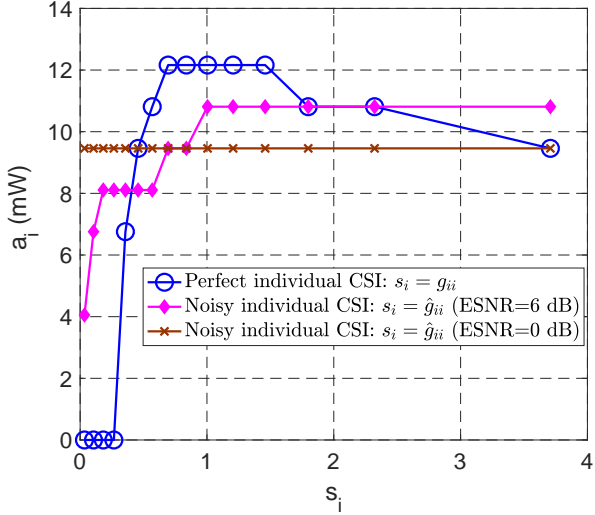}
    \end{center}
  \caption{The cooling effect observed for the previous figure is confirmed when considering other definitions for the energy-efficiency-based utility function (here with $\psi(x) = \log_2(1+x)$ and $P_0=10$ mW). In fact, it may also occur even in the absence of estimation noise.}
   \label{fig:dec_fun2}
\end{figure}

Fig.~\ref{fig:dec_fun} represents the decision function $\overline{f}_i(s_i)$ provided by Algorithm 1 for various values of ESNR while maximizing the sum-energy-efficiency, with equal weights for individual utilities, i.e. $\forall i \in \{1,...,K\}, \lambda_i = \frac{1}{K}$.  For this figure we assume that $\psi(x) = e^{-\frac{c}{x}}$, $c=1$, and $P_0=0$. At least two very interesting practical insights can be extracted from the figure. First, when perfect individual CSI is available (i.e., when $\mathrm{ESNR} \rightarrow \infty$), the optimal decision function naturally exhibits a threshold below which the transmitter should not transmit. This is very interesting since, to our knowledge, no paper on energy-efficient power control (at least in the sense as defined as in the present paper) has exhibited the need for a threshold, and this, in the absence of QoS constraints. This also allows one to make an interesting connection with \cite{paul-icc} where the sum-rate maximization is obtained with a thresholding technique and by merely using binary power control; our results show that this is more general and applies to other utility functions. Second, we see that, when only noisy estimation is available for the direct channel gain $g_{ii}$, the optimal decision functions comprises some piecewise constant parts. This shows, in particular, that not all available transmit power levels are exploited to maximize the average sum-energy-efficiency. Indeed, here $15$ levels are available but it is seen e.g., that when $\mathrm{ESNR} =6$ dB only some of them are exploited on the plot represented here. This situation may be referred to as a \textit{``cooling effect''} since less and less power levels are exploited as the $\mathrm{ESNR}$ decreases (in connection with the literature of learning when the chosen action consists of a tradeoff between exploration and exploitation -see e.g., \cite{lasaulce-book-2011}). Indeed, when the estimation noise is stronger ($\mathrm{ESNR} =0$ dB) this cooling effect on the decision function is completely apparent since only one transmit power level is exploited. In this respect, Fig.~\ref{fig:dec_fun2} represents the decision functions provided by Algorithm 1 when $\psi(x) = \log_2(1+x)$ and $P_0=10$ mW (as chosen in \cite{Zappone-TWC-2013}). {Interestingly, the cooling effect appears again and even in the absence of estimation noise, showing it is strongly related to the utility function form. One of the key messages our study conveys is that the best global performance may still be obtained even after reducing the possible choices in terms of transmit power levels. Having reduced action spaces may be very attractive in terms of computational complexity but also for measuring or sensing accurately the activity of the transmitters of interest.}


\subsection{Influence of the available CSI on the achievable long-term utility region}
 
Another type of precious information which is currently not available in the literature is the utility region for the problem of power control under partial information. Indeed, the knowledge of the long-term utility region is instrumental since it allows the best performance of the system to be fully characterized. In particular, any proposed power control scheme can be represented on the utility region and therefore, assessed in terms of efficiency. Again, for ease of exposition, we assume two transmitter-receiver pairs ($K=2$), which means that the utility region can be represented in a plane but the proposed framework is valid for any value of $K\geq2$. {Additionally, we choose $10 \log_{10}\left( \frac{\mathbb{E}(g_{ii})}{\mathbb{E}(g_{ji})}   \right) =0$ dB ($j\neq i$) to better illustrate the influence of the different channels.  The other system configurations are the same as Sec.~V-A.}

Fig.~\ref{fig:EEregion} represents the Pareto-frontier of the long-term utility region (thus in the $(U_1,U_2)$ plane) for the same scenarios as in the previous subsection. This allows one to see the impact of the individual channel gain quality in terms of achievable utility. The outer curve is obtained by assuming perfect individual CSI at the transmitters ($s_i = g_{ii}$) and using exhaustive search, which means that the curve represents exactly the best performance achievable under the considered partial information; interestingly, almost the same curve has been obtained by using Algorithm 1, which indicates that the corresponding optimality loss is negligible here. The other curves are obtained with the different values of ESNR considered for Fig.~\ref{fig:dec_fun} (namely, $\mathrm{ESNR}\in \{0, 6, +\infty\}$ dB) and using Algorithm 1. This result brings new insights w.r.t. existing works since it allows one to quantify the impact of the channel estimation quality on the final performance of the resource allocation policy i.e., when measured in terms of energy-efficiency. In the scenario, the cost of imperfect knowledge in terms of individual CSI is seen to be approximately $25 \% $ when the estimates are very noisy (namely $\mathrm{ESNR}=0$ dB).

From here on, we assume no estimation noise and rather assess the influence of partial information in terms of what channel gain is known. We define four information scenarios: 1. Perfect global CSI\footnote{This is generally considered as a very demanding assumption. Interestingly, some works show that it might be acquired in realistic settings. For instance, in \cite{chao-twc}, it is proved by having SINR feedback at the transmitter and using the idea of embedding local CSI into the transmit power levels, global CSI can be recovered at every transmitter. Another possibility to acquire CSI is to assume direct communication links between the transmitters, making the exchange of local CSI possible.}: $s_i = (g_{11}, g_{12}, g_{21}, g_{22})$; 2. Perfect direct CSI $s_i = (g_{11}, g_{22})$; 3. Perfect local CSI $s_i = (g_{ii}, g_{ji}), j \neq i$; 4. Perfect individual CSI: $s_i = g_{ii}$. Fig.~\ref{fig:EEregion} represents the Pareto-frontier of the long-term utility region (always in the $(U_1,U_2)$ plane) for these four scenarios.  Note that local CSI may be acquired e.g., when each transmitter sends a training sequence which is known to all the receivers. Then, Receiver $i$ can estimate $(g_{1i},\dots,g_{Ki})$ and feedback it to Transmitter $i$. Several useful observations can be made. First, moving from individual to local CSI does only bring a very marginal improvement in terms of energy-efficiency. On the other hand, knowing all the direct channels is definitely very useful for reaching good global performance. Third, the loss induced by not having global CSI is clearly assessed here and might be found to be acceptable. However, note that for the sake of representation, we assume two transmitter-receiver pairs here. For more users, these conclusions would need to be refined. Further, we provide simulations for a more typical number of users and assess the performance under these conditions.

\begin{figure}[h]
   \begin{center}
        \includegraphics[width=.48\textwidth]{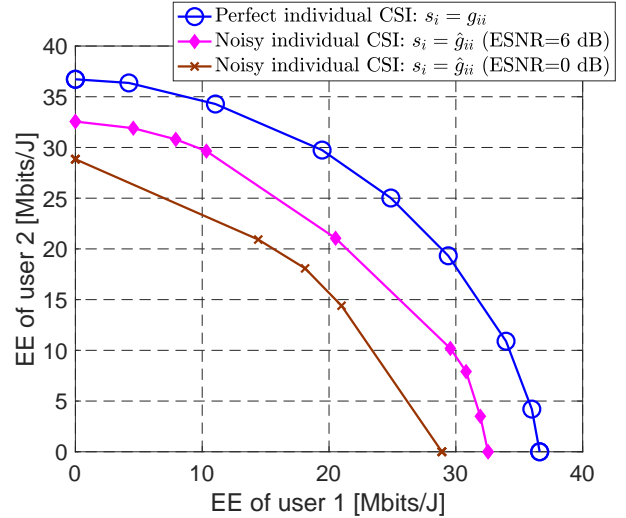}
    \end{center}
  \caption{Interestingly, the loss induced by having noisy individual channel gain estimates instead of perfect estimates is seen to be reasonable even when the estimates are very noisy.}
   \label{fig:EEregion}
\end{figure}

\begin{figure}[h]
   \begin{center}
        \includegraphics[width=.48\textwidth]{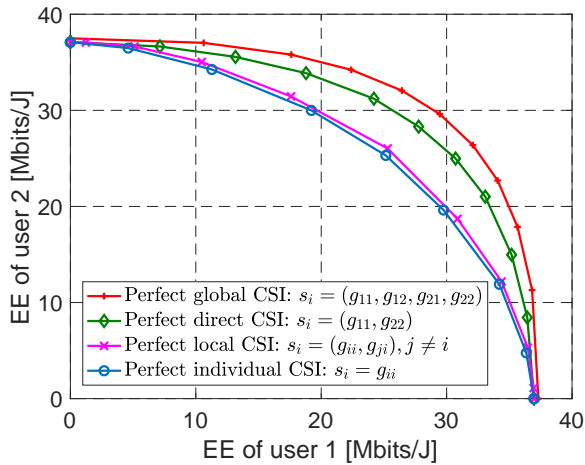}
    \end{center}
  \caption{Here, the knowledge is assumed to be partial but perfect. While knowing the cross channel gains is seen to bring very marginal improvement, the knowledge of direct channels allows one to bridge a quite good fraction of the gap between the individual CSI and global CSI scenarios.}
   \label{fig:EEregion}
\end{figure}

\subsection{Comparison between Algorithm 1 and the state-of-the-art}
\label{sec:comparison} 
 
Here (see Fig.~5), we consider a more general and generic wireless scenario namely, a small cell network (see e.g. \cite{chao-twc}\cite{Samarakoon-2016}) for which the interaction between $K=9$ neighboring cells is studied. The communication scenario is the same as the one considered in \cite{chao-twc}. In this model, the path loss effects for the link $ij$ are denoted by $\mathbb{E}(g_{ij}^s)$ and is inversely proportional to the distance between Transmitter $i$ and Receiver $j$. More precisely, $\mathbb{E}(g_{ij}^s) = \left(\frac{d_0}{d_{ij}}\right)^2$ where $d_{ij}$ denotes the aforementioned distance and $d_0= 5$ m is a normalization factor. All small base stations are considered to be in the center of their own cells, whereas the mobile stations $\mathrm{MS}_1, ..., \mathrm{MS}_9$ have been chosen  to have the following normalized coordinates : $(3.8, 3.2)$, $(7.9, 1.4)$, $(10.2, 0.7)$, $(2.3, 5.9)$, $(6.6, 5.9)$, $(14.1, 9.3)$, $(1.8, 10.6)$, $(7.1, 14.6)$, $(12.5, 10.7)$. One can obtain the real coordinates for the mobile stations by scaling these coordinates by a multiplying factor of $\frac{\mathrm{ISD}}{d_0}$, where ISD denotes the inter-site distance. The choices made for the parameter values are almost the same as in \cite{Zappone-TWC-2013} except for the SNR which is set here to a value that is more suitable for small cell networks (namely, $\mathrm{SNR}=$ 30 dB).  More precisely: $K=9$, $S=1$, $P_{\max}=10$ dBW, $\sigma^2=10 $ dBm, $P_0=10$ dBm, $| \mathcal{P}|=2000$ with uniform power increment, $| \mathcal{G}|=6$.

We compare the performance of Algorithm 1 to three relevant energy-efficient power control policies. The first is given by the non-cooperative power control policy of \cite{goodman-2000}, which is the first energy-efficient policy and is being reused in many papers in various settings (e.g., for multi-band MACs \cite{meshkati-poor}, for MIMO channels \cite{Bacci-TSP-2013}\cite{Veronica2011}, for channels with relays \cite{Zappone-TWC-2013}). The second scheme (called cooperative power control in \cite{Zappone-TWC-2013}) corresponds, to our knowledge, to the best way to maximize global energy-efficiency by imposing the available observation to be the individual CSI. The third scheme relies on pricing and has been introduced in \cite{goodman}; in contrast with two former policies, the latter policy assumes perfect SINR feedback and that for each data block several SINR measurements are available to iteratively update the power till convergence. Although the comparison with the latter scheme is not perfectly rigorous (in fact, it can be proved that SINR feedback may be sufficient to recover global CSI at every transmitter \cite{chao-twc}), we wanted here to consider this type of information since it is very popular in the literature of power control. In Fig.~\ref{fig:Zapp-vs-us-goodman} and Fig.~\ref{fig:Zapp-vs-us-log}, for both choices of $\psi$, the gain brought by using Algorithm 1 is seen to be very significant. Additionally, note that most of the existing power control schemes  have not been designed to deal with noisy estimates or an arbitrary partial knowledge about the global CSI, as opposed to Algorithm 1 that can always be used even in these complex scenarios. 

\begin{figure}[h]\label{fig:small-cells-network}
\begin{center}
\large
\begin{tikzpicture}[
    xscale = .52,yscale=.52, transform shape, thick,
    grow = down,  
    level 1/.style = {sibling distance=3cm},
    level 2/.style = {sibling distance=4cm}, 
    level 3/.style = {sibling distance=2cm}, 
    level distance = 1.25cm
  ]
 \draw [dotted] (10,0) -- (10,15);
\draw [dotted] (0,5) -- (15,5);
\draw [dotted] (0,10) -- (15,10); 
\draw (0,0) -- (15,0) -- (15,15) -- (0,15) -- (0,0);
\draw [dotted] (5,0) -- (5,15);

\node (b1) at (2.5,2.5)[shape = circle,draw]{SBS$_1$};
\node (b2) at (7.5,2.5)[shape = circle,draw]{SBS$_2$};
\node (b3) at (12.5,2.5)[shape = circle,draw]{SBS$_3$};
\node (b4) at (2.5,7.5)[shape = circle,draw]{SBS$_4$};
\node (b5) at (7.5,7.5)[shape = circle,draw]{SBS$_5$};
\node (b6) at (12.5,7.5)[shape = circle,draw]{SBS$_6$};
\node (b7) at (2.5,12.5)[shape = circle,draw]{SBS$_7$};
\node (b8) at (7.5,12.5)[shape = circle,draw]{SBS$_8$};
\node (b9) at (12.5,12.5)[shape = circle,draw]{SBS$_9$};
      

\node (m1) at (3.91101, 3.6508) [shape = rectangle,draw,fill=gray!10]{MS$_1$};
\node (m2) at (7.9828 ,1.01402) [shape = rectangle,draw,fill=gray!10]{MS$_2$};
\node (m3) at (10.1921 ,0.7040) [shape = rectangle,draw,fill=gray!10]{MS$_3$};
\node(m4) at (2.2934 ,5.8828) [shape = rectangle,draw,fill=gray!10]{MS$_4$};
\node(m5) at (6.6272 ,5.9110) [shape = rectangle,draw,fill=gray!10]{MS$_5$};
\node(m6) at (14.1351 ,9.2577) [shape = rectangle,draw,fill=gray!10]{MS$_6$};
\node(m7) at (1.8307 ,10.5872) [shape = rectangle,draw,fill=gray!10]{MS$_7$};
\node(m8) at (7.1415 ,14.6032) [shape = rectangle,draw,fill=gray!10]{MS$_8$};
\node(m9) at (12.5028 ,10.6665) [shape = rectangle,draw,fill=gray!10]{MS$_9$};

  \draw [->] (b9) to node[rectangle,left]{} (m9);
  \draw [->] (b1) to node[rectangle,left]{} (m1);
  \draw [->] (b2) to node[rectangle,left]{} (m2);
  \draw [->] (b3) to node[rectangle,left]{} (m3);
  \draw [->] (b4) to node[rectangle,left]{} (m4);
  \draw [->] (b5) to node[rectangle,left]{} (m5);
  \draw [->] (b6) to node[rectangle,left]{} (m6);
  \draw [->] (b7) to node[rectangle,left]{} (m7);
  \draw [->] (b8) to node[rectangle,left]{} (m8);

\draw [<->,color=blue,double] (10.2,14.8) to node[rectangle,fill=gray!1,below,yshift=-.3cm,xshift=.7cm]{Cell size: $d\times d$} (14.5,14.8);

\draw [<->,color=blue,double] (14.8,10.2) to node[rectangle]{} (14.8,14.5);

\draw [<->,double,color=blue] (2.5,11.5) to node[rectangle,fill=gray!1,below,yshift=-.3cm]{Inter-site distance: $d$} (7.5,11.5);
\draw [<->,double,color=blue] (8.5,12.5) to node[rectangle,fill=gray!1,right,xshift=.1cm]{$d$} (8.5,7.5);

%
  \draw [->,dashed,color=red] (b7) to node[rectangle,above,yshift=.3cm,xshift=-.3cm,fill=gray!1]{Interference} (m8);

%
%
%
%

\end{tikzpicture}
\caption{Assumed small cell scenario for Sec.~\ref{sec:comparison}.}
\end{center}
\label{fig:scn}
\end{figure}
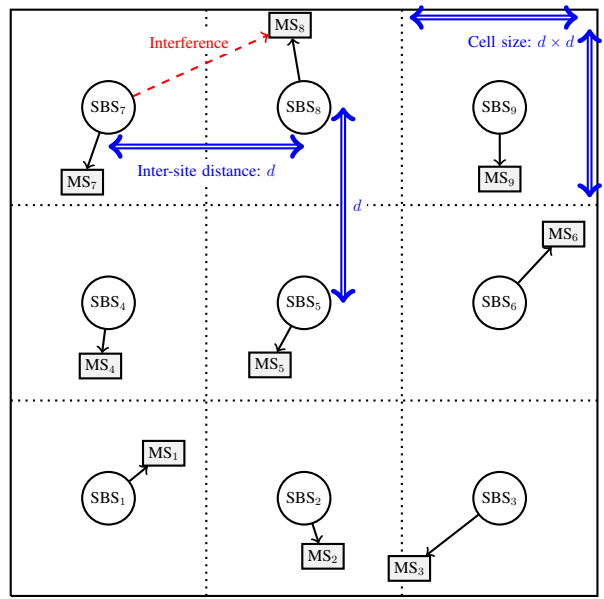

\begin{figure}[h]
   \begin{center}
        \includegraphics[width=.48\textwidth]{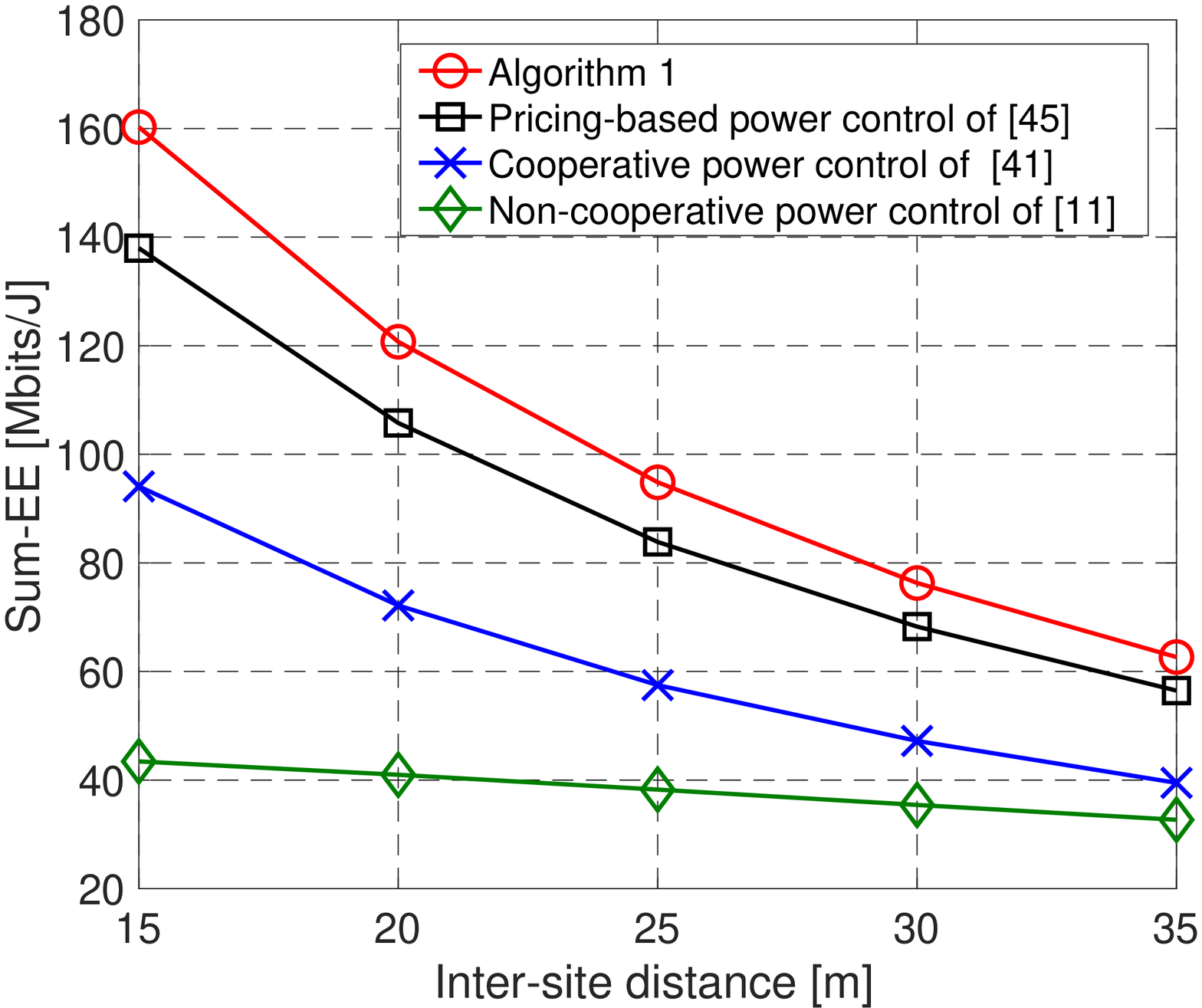}
    \end{center}
  \caption{Comparison of Algorithm 1 with state-of-the-art power control schemes for $\psi(x) = 1 - e^{-x}$ for the scenario of Fig.~5 (namely, with $K=9$ users.)}
   \label{fig:Zapp-vs-us-goodman}
\end{figure}
\begin{figure}[h]
   \begin{center}
        \includegraphics[width=.48\textwidth]{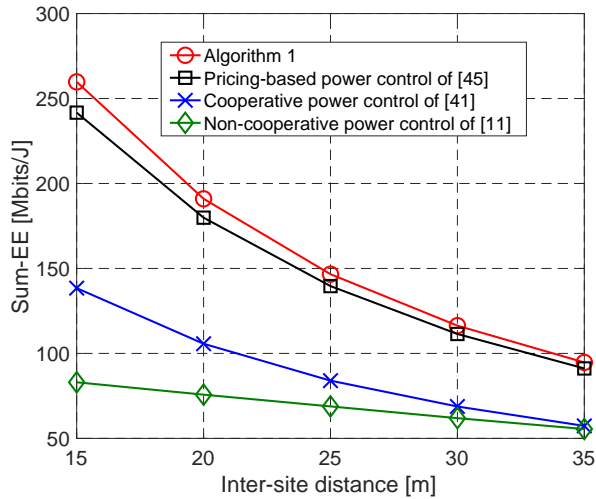}
    \end{center}
  \caption{Comparison of Algorithm 1 with state-of-the-art power control schemes for $\psi(x) =\log_2(1+x)$ for the scenario of Fig.~5 (namely, with $K=9$ users.)}
   \label{fig:Zapp-vs-us-log}
\end{figure}

\subsection{Additional simulations}

\begin{figure}[h]
   \begin{center}
        \includegraphics[width=.48\textwidth,height=.5\textwidth]{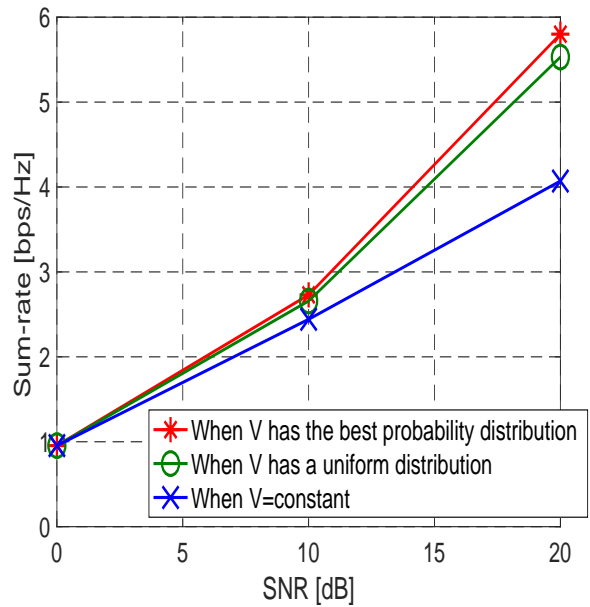}
    \end{center}
  \caption{Considering the sum-rate utility for $K=2$ with quality of service constraints, it is seen that exchanging a coordination key offline (and built from a lottery given by the random auxiliary variable $V$) brings a non-negligible improvement, especially at high SNR and regarding the underlying ease of implementation.}
   \label{fig:aux}
\end{figure}

To conclude the simulation section, we study a simple scenario with QoS constraints to show how the influence of the auxiliary variable on the global performance. A multiple access channel (MAC) with two transmitters is assumed ($K=2$); this is a special case of the general interference channel scenario studied so far. The channel chain of the link between Transmitter $i$ and the receiver is denoted by $g_i$. The instantaneous utility function for Transmitter $i$ is chosen to be:
\begin{equation}
u_i^{\mathrm{MAC}}(a_1,a_2;g_1,g_2)= \log_2\left(1+\frac{g_i a_i}{\sigma^2+ g_{-i} a_{-i}}\right) 
\end{equation}
with: the notation $-i$ stands for the other transmitter; $a_i \in \{0,P_{\max}\}$, $g_{i} \in \{0.3,1\}$; $\Pr(g_i=0.3)=\Pr(g_i=1)=50\%$ ; $s_i=g_i$; $\mc{V}=\{v_1, v_2 \}$. The goal is to maximize the long-term sum-utility under individual QoS constraints: $\mathbb{E}[u_1^{\mathrm{MAC}}] \geq 0.45\times\displaystyle\log_2(1+\mathrm{SNR})$ and $\mathbb{E}[u_2^{\mathrm{MAC}}] \geq 0.15\times\displaystyle\log_2(1+\mathrm{SNR})$ with $\mathrm{SNR}=\frac{P_{\max}}{\sigma^2}$. Fig.~\ref{fig:aux} represents the long-term sum-rate against SNR for three different cases: without auxiliary variable (bottom curve), with a uniformly distributed auxiliary variable (middle curve), and with the optimally distributed auxiliary variable (top curve).  The gain with respect to the case without auxiliary variable (namely, $V=const$) is seen to be appreciable. The optimal probability distributions $P_{A_i|S_i,V}$ and $P_V$ can be obtained by maximizing $W_{\lambda}$ ($\lambda_1=\lambda_2=50\%$) under QoS constraints. For instance, when $\mathrm{SNR}=20\mathrm{dB}$, an optimal distribution for $V$ is given by  $P_V(v_1)=51.6\%$ (when $V=v1$,  only Transmitter 1 transmits with $P_{\max}$,  Transmitter 2 keeps silent) and $P_V(v_2)=48.4\%$ (when $V=v2$, only Transmitter 2 transmits with $P_{\max}$, Transmitter 1 keeps silent). This clearly shows the benefit from using the auxiliary variable even in practice. This is very interesting since implementing a corresponding coordination mechanism is easy. For instance, it would consist in exchanging offline a sequence of binary realizations of a Bernouilli variable, which is perfectly doable when designing a real wireless system. 

\begin{figure}[h]
   \begin{center}
        \includegraphics[width=.48\textwidth]{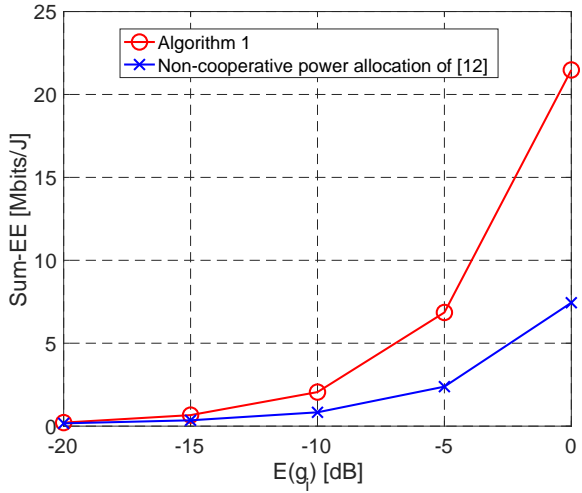}
    \end{center}
  \caption{For multi-band MAC and typical values for the channel gain mean ($\mathbb{E}(g_i) \geq 0.1$), the proposed power control scheme is shown to provide a significant performance gain over the technique proposed in \cite{meshkati-poor}.}
   \label{fig:PMAC}
\end{figure}

\begin{figure}[h]
   \begin{center}
        \includegraphics[width=.48\textwidth]{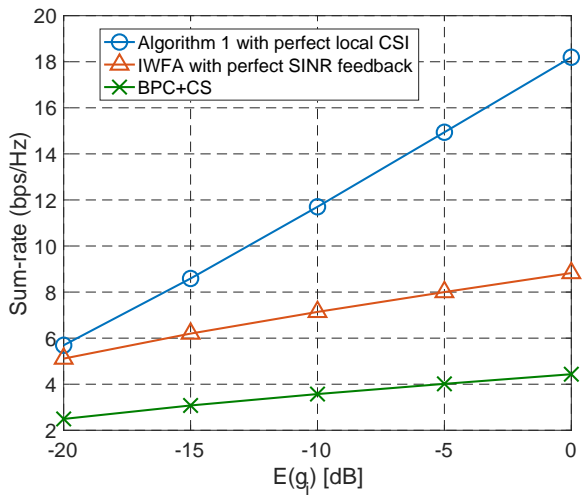}
    \end{center}
  \caption{Considering the sum-rate in multi-band MAC and typical values for the channel gain mean ($\mathbb{E}(g_i) \geq 0.1$) scenario, the proposed power control scheme is shown to provide a significant performance gain over the technique in the litterature.}
   \label{fig:PMAC_SR}
\end{figure}

Another interesting case to consider is the multi-band MAC scenario, which has been treated in \cite{meshkati-poor}. We consider the sum-energy-efficiency utility and compare it with the scheme derived in \cite{meshkati-poor}. For this, we assume $K=3$ users, $2$ possible operating bands, and $\psi(x) = (1-e^{-x})^{100}$. The other parameters are: $P_{\max}=10$ dBW, $\sigma^2=10 $ dBm, $P_0=10$ dBm, $| \mathcal{P}|=2000$ with uniform power increment, $| \mathcal{G}|=6$. Fig.~\ref{fig:PMAC} shows the sum-EE performance against the average link quality $\mathbb{E}(g_i)$ for the scheme obtained from Algorithm 1 and the scheme proposed in \cite{meshkati-poor}. The performance gain is appreciable even for a quite small ratio $\frac{K}{B}$; indeed, the sum-utility performance is improved by more than three times. Moreover, it  has been observed for other simulations that the gain is even more significant when the load per band increases, e.g., in the scenario $K=9$, $B=2$. Fig.~\ref{fig:PMAC_SR} depicts the sum-rate against the average link quality $\mathbb{E}(g_i)$ for the different schemes obtained from: Algorithm 1, the well known iterative water-filling algorithm (IWFA), and the binary power control (BPC) plus channel selection (CS) scheme. The parameters chosen are same as those in Fig.~9. The sum-rate is defined as
\begin{equation}
u^{\mathrm{MAC}}= \sum_i\log_2\left(1+\frac{g_i a_i}{\sigma^2+ \sum_{j\neq i}g_{j} a_{j}}\right). 
\end{equation}
IWFA aims at maximizing the individual transmission rate at each transmitter exploiting the SINR feedback. The BPC plus CS is based on the fact that BPC has been shown to be an efficient (and sometimes optimal) power control in one band MAC, and thus in multi-band case the transmitter can choose one channel (with the largest channel gain) first and transmit at full power over this channel. The performance gain is significant especially for large channel gain means, which can be explained by the fact that coordination becomes more useful as interference becomes stronger.

\section{Conclusion}
\label{sec:concl}

To summarize in a concise way what the proposed approach brings w.r.t. the state-of-the-art and what its limitations are, we propose to describe its strengths and weaknesses under a list form.

\textbf{Strong features of our approach}\\
$\blacktriangleright$ In contrast with the state-of-the-art, by making a fruitful connection between power control and information theory, our approach allows one to characterize the best performance a set of transmitters can achieve in terms of power control under partial information. In particular, this allows one to measure the efficiency of any proposed power control scheme.\\
$\blacktriangleright$ Both the limiting performance analysis and the proposed distributed algorithm work for a broad class of utility functions and not only for a specific utility function as often assumed in the literature. \\ 
$\blacktriangleright$ Both the limiting performance analysis and the proposed algorithm work for a broad class of partial observation structures and not only for a very specific observation structure as often assumed in the literature. For instance, the vast majority of power control and radio resource allocation schemes (see e.g., \cite{paul-icc}\cite{yu-2002}\cite{scutari-tsp-2009}\cite{Zappone-TWC-2013}\cite{mael-twc}\cite{larrson-beamforming}) makes information assumptions such as perfect individual or global CSI but does not allow one to deal with noisy estimation or other arbitrary partial and perfect information.\\
\textbf{Limitations of our approach}\\
$\blacktriangleright$ Although assuming the power control actions and network state to be discrete does not constitute a limitation for the limiting performance analysis since the continuous case follows by specialization (namely, as a limiting case of the discrete case as done for classical coding theorems), it typically involves some complexity limitations for the proposed algorithm. The proposed algorithm corresponds to one possible numerical solution to determine good power control functions, however finding low complexity numerical techniques constitutes a very relevant issue to be explored.\\ 
$\blacktriangleright$ Both the limiting performance analysis and the proposed algorithm assume utility functions under the form $u_i(a_0, a_1,...,a_K)$ when $a_0$ corresponds to the realizations of an i.i.d. random process $(A_{0,t})_{t\geq 1}$ and the partial information available to Transmitter $i$ (namely, $s_i$) is the output of a discrete memoryless channel. In this paper, the channel state is assumed to be i.i.d. which is a common and very well accepted assumption. If the state or the observation structure happens to be with memory, the derived results would need to be generalized. This would be necessary for instance, for a Markovian state.\\ 
$\blacktriangleright$ As many related papers, the proposed algorithm provides power control functions under a numerical form but not in an analytical form. However, the obtained numerical results may be used as a source of inspiration to propose relevant classes of functions which are suited to the considered setup. Thresholding, saturation, steps, scaling are examples of operations which may be exhibited and used.

\bibliographystyle{unsrt}

%
%

\clearpage
\newpage
\thispagestyle{empty}


\end{document}